\documentclass{article}
\usepackage{spconf,amsmath,graphicx}
\usepackage{amssymb}
\usepackage{tabularx}
\usepackage{multirow}
\usepackage{booktabs}
\usepackage{hyperref}

\title{FedSODA: Federated Cross-assessment and Dynamic Aggregation for Histopathology Segmentation}
\name{Yuan Zhang$^{1}$, Yaolei Qi$^{1}$, Xiaoming Qi$^{1}$, Lotfi Senhadji$^{2,3}$, Yongyue Wei$^{4}$, Feng Chen$^{5}$, Guanyu Yang$^{1,2 \dagger}$ \thanks{$\dagger$ Corresponding author: Guanyu Yang (email: yang.list@seu.edu.cn)
This research was supported by the Intergovernmental Cooperation Project of the National Key Research and Development Program of China (2022YFE0116700). We thank the Big Data Computing Center of Southeast University for providing the facility support.
}}
\address{$^{1}$ Key Laboratory of New Generation Artificial Intelligence Technology and Its Interdisciplinary\\ Applications (Southeast University), Ministry of Education
	$^{2}$Jiangsu Provincial Joint International\\ Research Laboratory of Medical Information Processing, Southeast University 
	$^{3}$Inserm, LTSI, UMR 1099,\\ University of Rennes,
	$^{4}$  Public Health and Epidemic Preparedness and Response Center, Peking University\\ 
	$^{5}$  Department of Biostatistics, School of Public Health, Nanjing Medical University
}
\begin{document}
%
\maketitle
\begin{abstract}
Federated learning (FL) for histopathology image segmentation involving multiple medical sites plays a crucial role in advancing the field of accurate disease diagnosis and treatment. However, it is still a task of great challenges due to the sample imbalance across clients and large data heterogeneity from disparate organs, variable segmentation tasks, and diverse distribution. Thus, we propose a novel FL approach for histopathology nuclei and tissue segmentation, FedSODA, via synthetic-driven cross-assessment operation (SO) and dynamic stratified-layer aggregation (DA). Our SO constructs a cross-assessment strategy to connect clients and mitigate the representation bias under sample imbalance. Our DA utilizes layer-wise interaction and dynamic aggregation to diminish heterogeneity and enhance generalization. The effectiveness of our FedSODA has been evaluated on the most extensive histopathology image segmentation dataset from 7 independent datasets. The code is available at   \href{https://github.com/yuanzhang7/FedSODA}{https://github.com/yuanzhang7/FedSODA} .
\end{abstract}
\section{Introduction}
\label{sec:intro}
Collaborative training across multiple sites for histopathology image segmentation, encompassing both nuclei and tissues, stands as a pivotal task in the field of precise disease diagnosis and treatment \cite{van2021deep}. Due to the intricate annotation of cellular boundaries and the imperative requirement to maintain privacy protections concerning clinical data across multiple sites, the consolidation of extensive labeled datasets to attain better model generalization remains restricted. Thus, an imperative arises for a more robust and generalized model.

\begin{figure}[!t]
	\centering
	\centerline{\includegraphics[width=\linewidth]{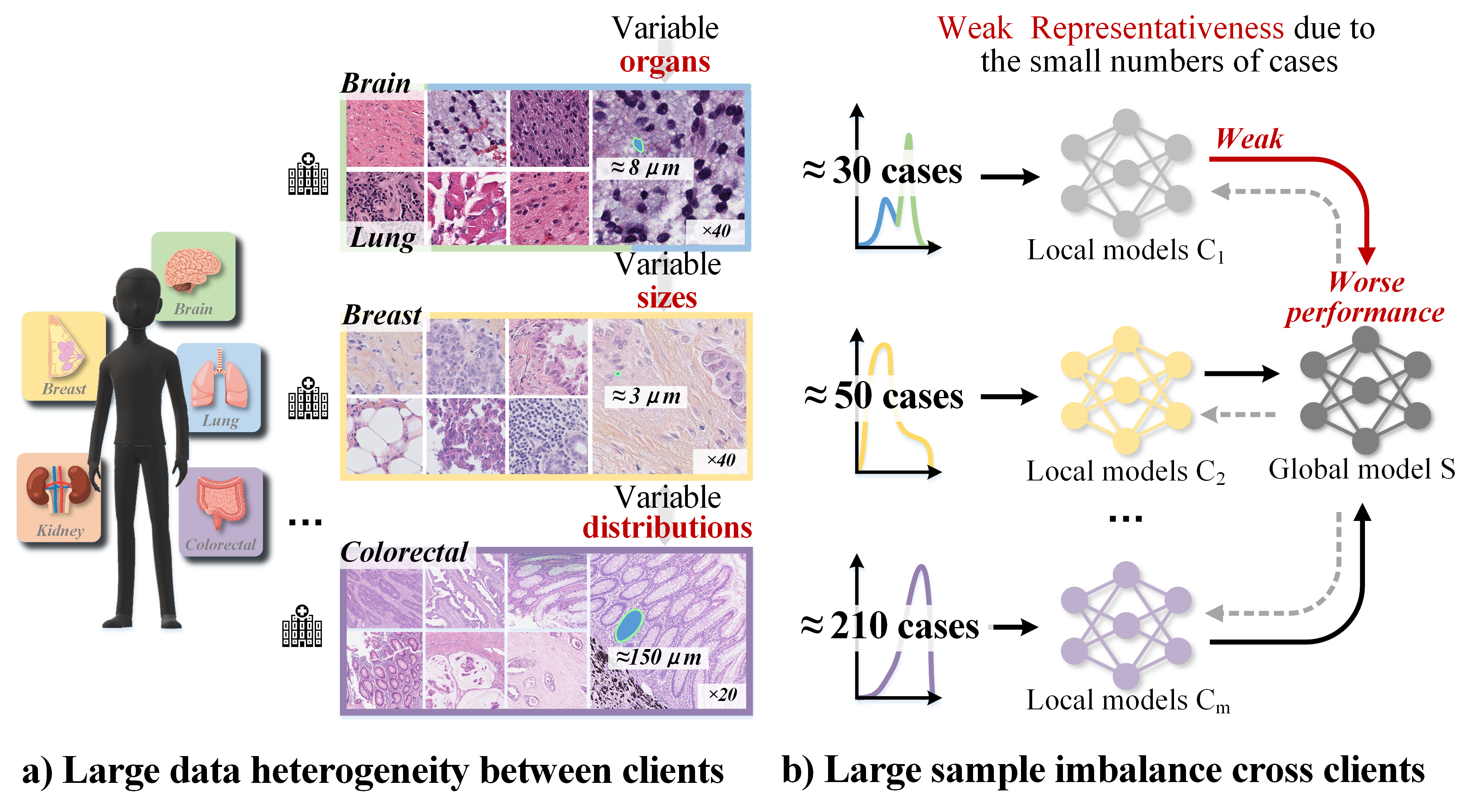}}
	\label{fig:1}
	\caption{\textbf{Challenges:} a) Large data heterogeneity arises from disparate organs, variable segmentation target sizes, and diverse intensity distributions. b) Sample imbalance from substantial variations in sample quantities across different clients.}
\end{figure}

However, there are two main challenges: 1)\textbf{ Data heterogeneity}. The huge variations in data distribution across different clients stem from the disparate organs, the varying sizes of segmentation targets, and the variable intensity derived from chromatin patterns. As shown in Figure~\ref{fig:1}. a), The size of cells and tissues from various organs exhibits a notable range, spanning from 3 $\mu$m to 150 $\mu$m. The significant discrepancy among clients further amplifies the disparity in features, leading to an inductive bias within the local models,  thereby manifesting as poor performance on the global model. 2) \textbf{Sample imbalance}. Due to various disease prevalences, sample imbalance arises in the data volume across clients, ranging from 30 to 210 (Figure~\ref{fig:1}. b)). Clients with limited samples are prone to over-fitting, resulting in weak performance on unseen data and an incapacity to attain accurate representation.

Fortunately, federated learning provides a promising paradigm, enabling joint learning across multiple centers without data sharing. The classic FedAvg \cite{McMahan2017fedavg} and its variants \cite{li2021model,qi2022contrastive} effectively establish mutually independent client training, with jointly optimized global models. Subsequently, noteworthy methods such as FedProx \cite{li2020fedProx}, FedBN \cite{li2021fedbn}, and HarmoFL \cite{jiang2022harmofl}, further explore the trade-offs between clients and the server in terms of the divergence among local models and the feature bias inherent in the global model. However, these prior works are not directly tailored to our specific task and struggle to address the above challenges in segmentation.

To tackle the above obstacles, we propose a novel approach called \textbf{FedSODA}, involving synthetic-driven cross-assessment operation (\textbf{SO}) and dynamic stratified-layer aggregation (\textbf{DA}). (1) For the challenge of sample imbalance, we propose a synthetic-driven cross-assessment operation to overcome the over-fitting of client models on unseen data. Our FedSODA synthesizes interaction information based on client distribution, constructs a cross-assessment strategy connecting clients, and mitigates the representation bias from clients with limited samples, thus improving the generalization and robustness of the global model. (2) For the challenge of data heterogeneity, we introduce dynamic stratified-layer aggregation to ease the inductive bias within local models due to size variations. Specifically, the shallow and deep layers present different feature representations \cite{guo2022isdnet} are conducive to segmenting small nuclei and large tissues respectively. Unlike weighting entire networks \cite{McMahan2017fedavg,li2020fedProx}, our FedSODA utilizes layer-wise interaction and autonomously updates suitable weights, thus emphasizing personalized representations for each client to better align with the size-specific features.

In summary, our FedSODA contributes in the following ways: (1) Synthetic-driven cross-assessment operation is proposed to construct a cross-assessment strategy connecting clients, thereby mitigating the representation bias under sample imbalance. (2) Dynamic stratified-layer aggregation is proposed to counteract the inductive bias within local models due to data heterogeneity, thus emphasizing personalized representations for each client to better align with the size-specific features. (3) To the best of our knowledge, our study marks the first instance of evaluating FL performance with the most extensive histopathology image segmentation dataset, derived from a compilation of seven independent datasets.

\section{Related Work}
\label{sec:format}
\subsection{Federated Learning}
Federated learning aims to coordinate clients through updates and communication, to acquire a global model at the aggregation server without leaking raw data. FedAvg \cite{McMahan2017fedavg} along with the subsequent works \cite{li2020fedProx, li2021fedbn, jiang2022harmofl} mentioned above, have propelled advancements on FL. However, these methods are not customized for our specialized histopathology image segmentation, rendering them less effective for the inherent challenges. FedProx \cite{li2020fedProx} focuses on global aggregation drift by adding a proximal term to each local objective. FedBN \cite{li2021fedbn} concerns feature shifts among clients by preserving local batch normalization parameters. Additionally, RHFL \cite{fang2022robust} aligns models by additional shared data to reduce negative effects from noisy clients. FSMAFL \cite{huang2022few} utilizes abundant public data to bridge the imbalance of limited private data. Nevertheless, the requirement for an additional and representative proxy dataset \cite{tan2022towards} raises the cost of data acquisition.

\subsection{Federated Learning for Histopathology Image}
In light of privacy concerns, the research community has begun exploring FL for histopathology images\cite{xia2023cross}. HistoFL\cite{Lu2022e} applies FL with randomized noise generation on gigapixel whole slide images. Prop-FFL\cite{hosseini2023proportionally} proposes an optimization objective function based on proportional fair resource allocation. ProxyFL\cite{kalra2023decentralized} suggests an extra proxy model to enhance privacy in prognosis, replacing the centralized server. However, these methods primarily focus on fairness or privacy issues without addressing the challenges of data heterogeneity and sample imbalance. Furthermore, these efforts have predominantly substantiated FL's effectiveness in classification and prognosis tasks, with limited attention given to segmentation\cite{He2022Seg,Qi_2023_ICCV}. HarmoFL\cite{jiang2022harmofl} firstly explores the local and global update drift by normalizing the frequency-space and amplitude for nuclei segmentation, which may not inherently tackle the challenge of heterogeneity between assorted segmentation targets. Different from existing works, we investigate a novel federated learning framework for sample imbalance with high heterogeneity in histopathology nuclei and tissue segmentation tasks, which dynamically aggregates clients to alleviate contradiction and enhance generalization.  

\begin{figure*}[t]
	\centering
	\centerline{\includegraphics[width=0.8\linewidth]{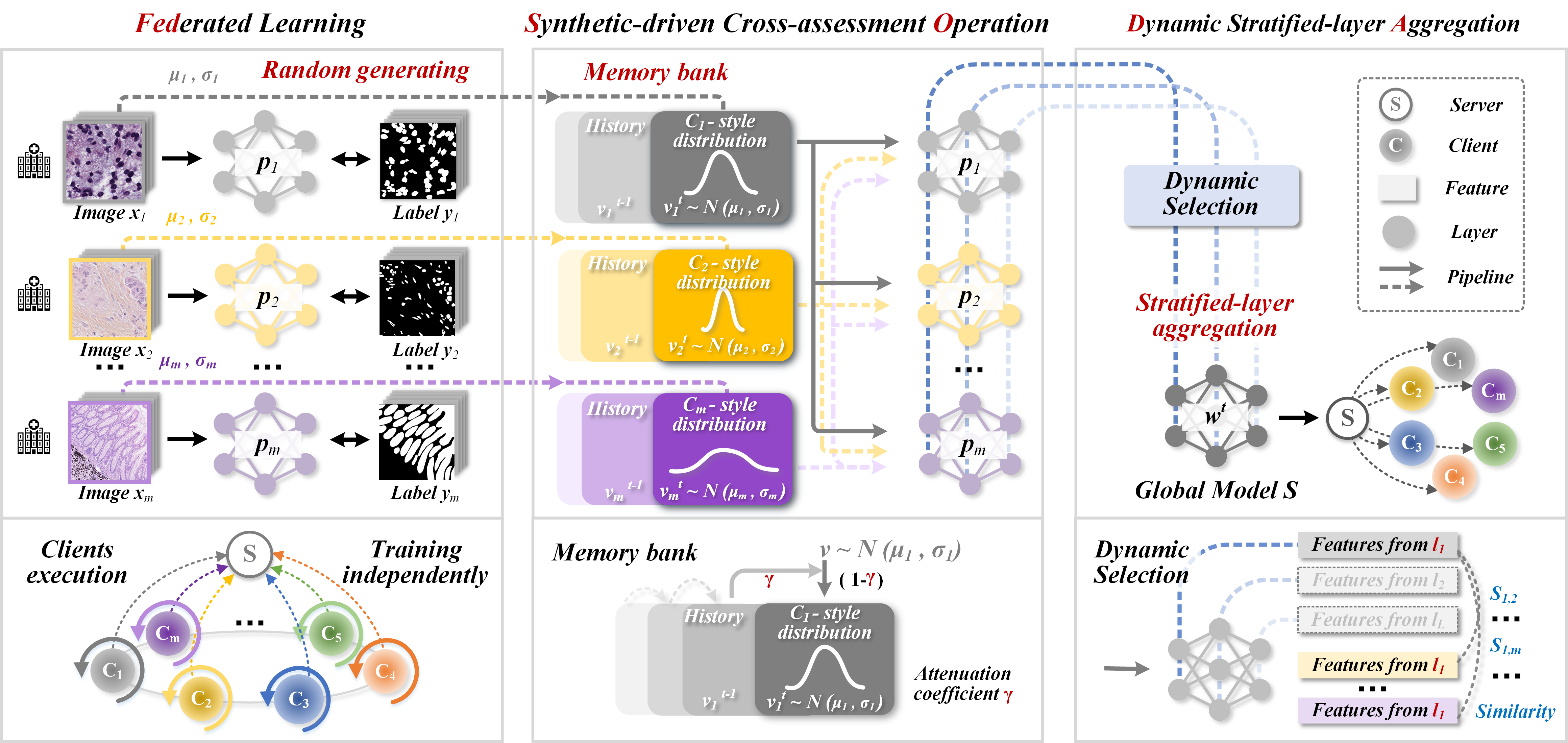}}
	\caption{\textbf{Methodology}. Schematic overview of our proposed FedSODA illustrated on the histopathology image segmentation. }
	\label{fig:2}
\end{figure*}

\section{Methodology}
\label{sec:format} 
\subsection{Synthetic-driven Cross-assessment Operation}
We propose a synthetic-driven cross-assessment operation to mitigate the sample imbalance between clients in Fig ~\ref{fig:2}. Each dataset has different distributions with mean $\mu$ and standard deviation $\delta$. In the $t_{th}$ communication, Gaussian synthetic information ${v^t} \sim\mathcal{N} (\mu,\sigma)$ is randomly generated and then uploaded to the server to evaluate the similarity between clients. Also, a memory bank is constructed to enhance the stability and consistency of the synthesized data space by combining the historical information from the previous communication rounds and newly generated samples from the current:
$ {v^{t}} =\gamma {v^{t-1}} +(1-\gamma) {v^{t}} $
where $\gamma$ is an attenuation coefficient used to control the proportion of historical information.
A segmentation consistency loss $\mathcal{L}_{sc}$ is introduced to further address the sample imbalance, promoting better coordination between clients. The consistency parameter is defined as $\xi= max(0, \lvert y-\hat{y} \rvert 
- \epsilon)$ with label-output pairs $\{y,\hat{y}\}$. The detailed formula is described as: $\mathcal{L}_{sc}=\mathcal{L}_{ce}(\xi y, \hat{y})$.

Finally, the segmentation accuracy is constrained by the combined effect of the segmentation consistency loss $\mathcal{L}_{sc}$  and cross-entropy loss $\mathcal{L}_{ce}$  functions. Our segmentation consistency loss $\mathcal{L}_{sc}$ penalizes incorrectly segmented boundary pixels prevents the segmentation of large organs from being overly influenced by small-sized cells and enhances optimization consistency across global models.

\subsection{Dynamic Stratified-layer Aggregation}
We propose a dynamic stratified-layer aggregation strategy to adaptively integrate shallow and deep layers of weights to enhance size-specific representation under data heterogeneity. The entire client network is stratified into $L$ layers with varying feature representation capabilities. 

The synthetic information $v_k^t$ generated based on the client $C_k$ is input into each client's parametric model on the server side to obtain the corresponding output features set $\{\mathbb{F}_{k, 1}, \mathbb{F}_{k, 2}, \ldots, \mathbb{F}_{k, m}\}$, where $m$ denotes the number of clients and $\mathbb{F}$ is the set of output features for each layer: $\mathbb{F} = \{f^1, f^2, \ldots, f^L \}$. Each stratified layer, such as layer $l$, performs cosine similarity calculations $cos(\cdot)$ cross clients to obtain the dynamic weights: $\{s^l_{k, 1}, s^l_{k, 2}, \ldots, s^l_{k, m} \}$, where $s^l_{k, m} = cos(f^l_{k, k}, f^l_{k, m})$. Then the normalization is performed as $\hat{s}^l_{k, m} = s^l_{k, m} / \sum {s^l}$. Then, the weights of the server model are dynamically aggregated layer by layer based on the interaction between clients. The $t$ communication round of the server model $w^{t,l}$ aggregates the $l$ layer of the $t-1$ round of client models with corresponding interaction weights:
\begin{align}
	{w^{t,l}}=\frac{1}{m}\sum_{k=0}^m [\lambda {p_k^{t-1}}+(1-\lambda)\sum_{j\neq k} {\hat{s}^l_{k,j} \cdot {p_j^{t-1}} }]
\end{align}
The updated server model $w^{t}$ is obtained through the above dynamic aggregation process across all layers $w^{t,l}$ and then distributed to each client model for the subsequent epoch of training: $ p^{t} \leftarrow{} w^{t}$. Our strategy grants the server model a heightened degree of interactive freedom, thereby enhancing the network's generalization performance.

\begin{table*}[!htb]
	\begin{center}
		\caption{The segmentation results of Dice and accuracy for seven clients. Each column represents one client.}
		\resizebox{\textwidth}{!}{
			\setlength\tabcolsep{3.5pt}
			\vspace{20pt}
			\begin{tabular}{ccccccccc|ccccccccc}
				\hline
				\textbf{} & \multicolumn{8}{c|}{\textbf{Dice \%}} &  & \multicolumn{8}{c}{\textbf{Accuracy \%}} \\ \cline{2-9} \cline{11-18} 
				& CoNsep & CPM-17 & CRAG & CryoNuSeg & Glas & Kumar & TNBC & Average &  & CoNsep & CPM-17 & CRAG & CryoNuSeg & Glas & Kumar & TNBC & Average \\ \hline
				\textbf{FedAvg} & 80.48 & \underline{87.01} & \underline{84.09} & \underline{81.81} & 86.97 & 79.99 & 78.01 & 82.62 & \textbf{FedAvg} & 92.45 & \underline{94.85} & \underline{86.58} & \textbf{93.96} & 87.58 & 92.18 & 95.47 & 91.87 \\
				\cite{McMahan2017fedavg} & 0.04 & 0.03 & 0.08 & 0.02 & 0.08 & 0.03 & 0.06 & 0.05 & \cite{McMahan2017fedavg} & 0.03 & 0.02 & 0.08 & 0.02 & 0.07 & 0.03 & 0.03 & 0.04 \\
				\textbf{FedProx} & 79.82 & 86.79 & 83.37 & 80.89 & 87.28 & 79.58 & 77.73 & 82.21 & \textbf{FedProx} & 92.32 & 94.72 & 86.53 & 93.55 & 87.33 & 92.07 & 95.44 & 91.71 \\
				\cite{li2020fedProx} & 0.04 & 0.04 & 0.08 & 0.03 & 0.08 & 0.04 & 0.06 & 0.05 & \cite{li2020fedProx} & 0.02 & 0.02 & 0.07 & 0.02 & 0.08 & 0.04 & 0.03 & 0.04 \\
				\textbf{FedBN} & \textbf{81.46 }& 86.94 & 82.24 & 80.92 & \underline{87.63} & 79.99 & \textbf{79.93} & 82.73 & \textbf{FedBN} & \underline{92.50} & 94.83 & 85.88 & 93.33 & \underline{88.22} & 92.18 & \textbf{95.90} & 91.83 \\
				\cite{li2021fedbn} & 0.04 & 0.04 & 0.06 & 0.02 & 0.07 & 0.03 & 0.04 & 0.04 & \cite{li2021fedbn} & 0.03 & 0.02 & 0.07 & 0.02 & 0.06 & 0.03 & 0.02 & 0.04 \\
				\textbf{HarmoFL} & 79.69 & 86.65 & 80.33 & 81.12 & 86.54 & \textbf{80.88} & 78.38 & 81.94 & \textbf{HarmoFL} & 92.06 & 94.74 & 84.09 & 93.39 & 86.97 & \underline{92.48} & 95.57 & 91.33 \\
				\cite{jiang2022harmofl} & 0.04 & 0.03 & 0.08 & 0.04 & 0.08 & 0.02 & 0.06 & 0.05 & \cite{jiang2022harmofl}  & 0.03 & 0.02 & 0.08 & 0.02 & 0.08 & 0.03 & 0.03 & 0.04 \\
				\textbf{FedSODA} & \underline{81.10} & \textbf{87.06} & \textbf{86.49} & \textbf{82.61} & \textbf{87.98} & \underline{80.17} & \underline{78.43} & \textbf{83.41} & \textbf{FedSODA} & \textbf{92.75} & \textbf{94.91} & \textbf{89.36} & \underline{93.80} & \textbf{88.68} & \textbf{92.52} & \underline{95.70} & \textbf{92.53} \\
				(Ours) & 0.03 & 0.03 & 0.07 & 0.02 & 0.08 & 0.03 & 0.06 & 0.05 & (Ours) & 0.03 & 0.02 & 0.06 & 0.02 & 0.08 & 0.03 & 0.03 & 0.04 \\ \hline
			\end{tabular}
		}
		\label{tab:t1}
	\end{center}
\end{table*}

\section{Experiment}
\label{sec:pagestyle}
\subsection{Data Details and Experimental Settings}
To evaluate our FedSODA for multi-site histopathology image segmentation, we conduct comprehensive experiments on seven datasets, including (1) CoNSeP \cite{graham2019hover}, (2) CPM-17 \cite{vu2019cpm17}, (3) CRAG \cite{graham2019mild}, (4) CryoNuSeg \cite{mahbod2021cryonuseg}, (5) Glas \cite{SIRINUKUNWATTANA2017489glas}, (6) Kumar \cite{kumar2017dataset}, and (7) TNBC \cite{naylor2018tnbc}. CRAG \cite{graham2019mild} and Glas \cite{SIRINUKUNWATTANA2017489glas} are tissues with relatively large numbers of samples, while the other 5 datasets are nuclei with a small number of samples. Each dataset is allocated to each client with its original division. U-Net \cite{Ronneberger2015} is trained for 300 epochs with 5 local update epochs for each communication round. All comparative experiments adopt identical initialization and communication round settings. We use a batch size of 4, Adam optimizer with a learning rate of $1e^{-4}$, and momentum of 0.9 and 0.95. All experiments are performed on an NVIDIA GPU 3090 with 24 GB. The Dice and accuracy are used to evaluate the results. 

\subsection{Experimental Results}
Our FedSODA achieves the best segmentation results compared with the relevant FL approaches \cite{McMahan2017fedavg,li2020fedProx,li2021fedbn,jiang2022harmofl}.

\textbf{Quantitative Results.} With the synthetic-driven cross-assessment operation and dynamic stratified-layer aggregation strategy mitigating local and global drifts, our method consistently outperforms others, averaging a Dice of 83.41\% and an accuracy of 92.52\% in Table \ref{tab:t1}. The notable boost of FedSODA on the large tissue dataset of CRAG, 6.16\% Dice higher than HarmoFL and 2.78\% accuracy higher than FedAvg, shows effective mitigation of data imbalance bias. Besides, FedSODA steadily improves the client results without sacrificing model performance on small nuclei datasets.

\begin{figure}[!t]
	\centering
	\centerline{\includegraphics[width=\linewidth]{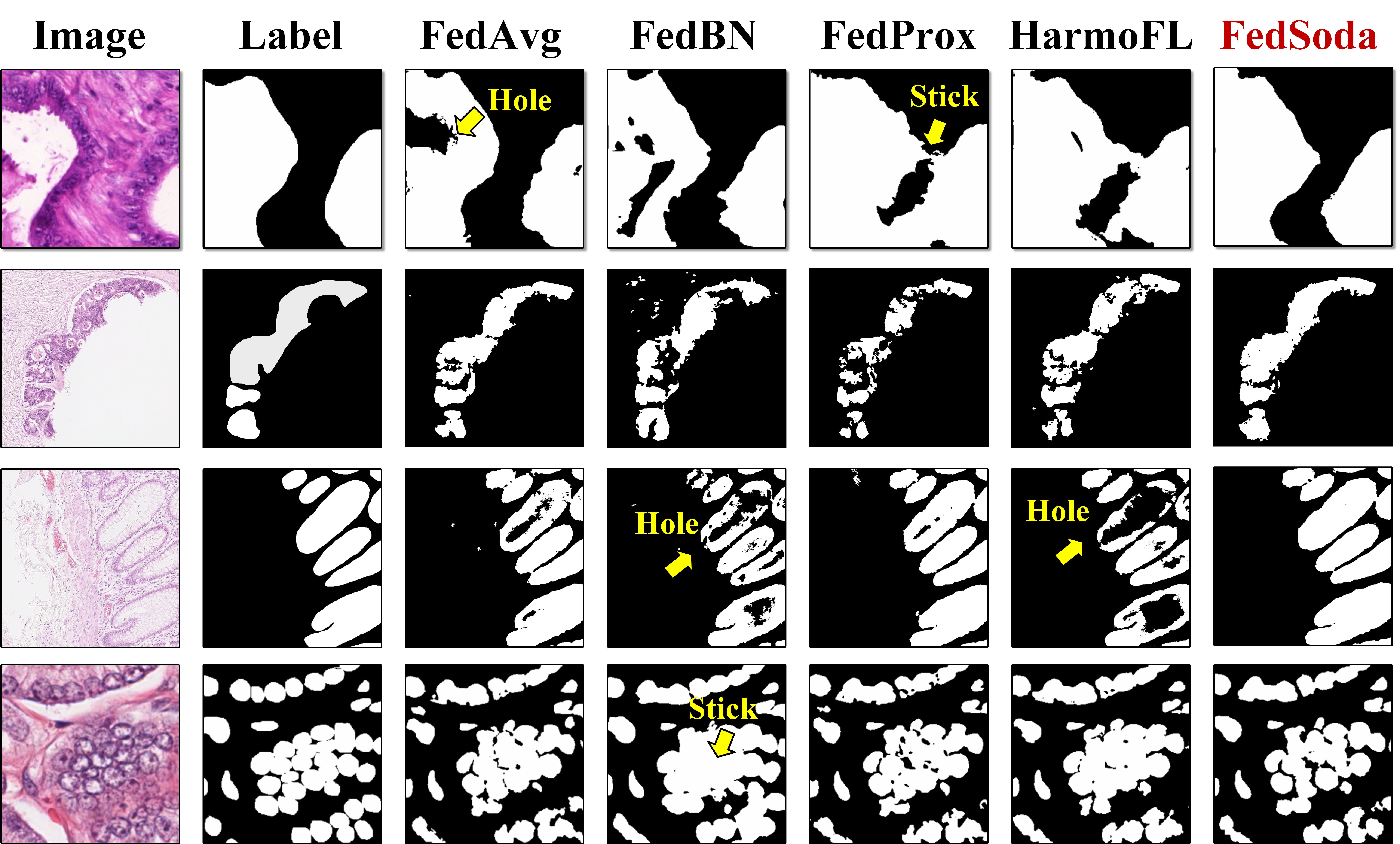}}
	\caption{The segmentation results for different models.}
	\label{fig:3}
\end{figure}

\textbf{Qualitative Results.} We visualize the segmentation results to demonstrate a qualitative comparison, as shown in Fig \ref{fig:3}. Our FedSODA reduces the amount of false positives and negatives and has a better shape and boundary segmentation output for histopathology nuclei and tissues. Compared with the second label column, other FL methods either contain more erroneous segmentation regions or unclear boundaries. With the proposed dynamic stratified strategy, our approach shows more accurate and smooth boundaries. Besides, visualization results from the first to third rows demonstrate superior FedSODA performance in completing holes and enhancing inter-structural consistency. The fourth row further shows our FedSODA better distinguishes clustered nuclei.

\begin{figure}[!t]
	\centering
	\centerline{\includegraphics[width=\linewidth]{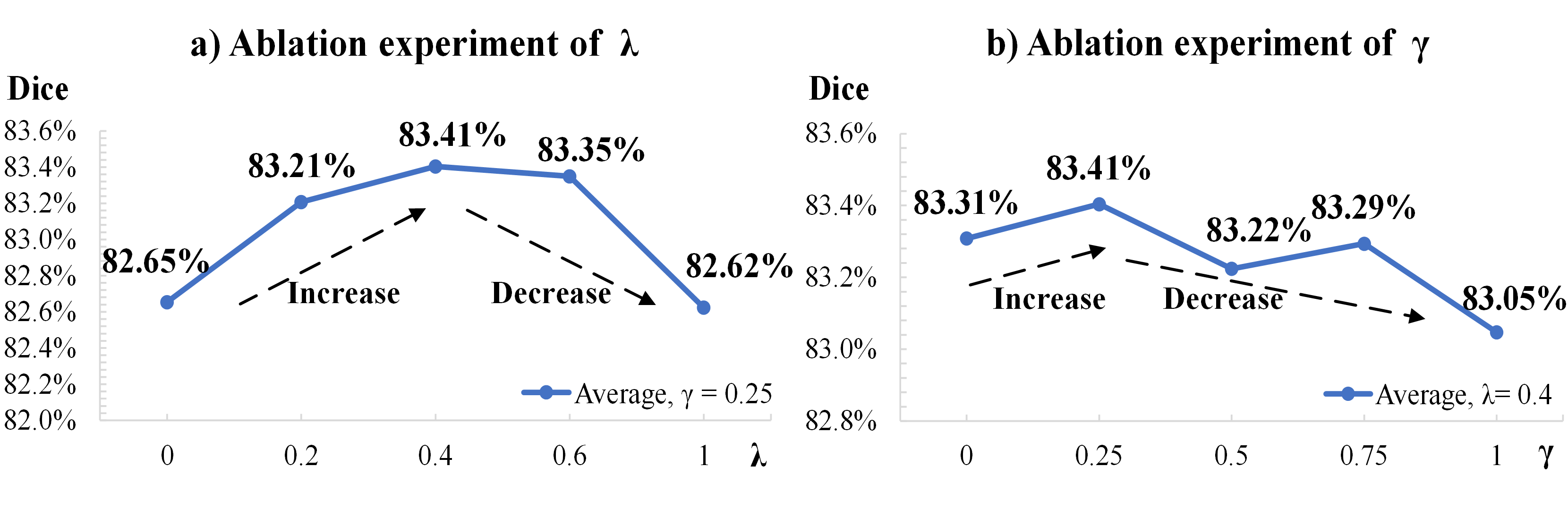}}
	\caption{Ablation analysis for hyper-parameters: a) Average Dice for different $\lambda$ with a fixed $\gamma=0.25$. b) Average Dice for different attenuation coefficient $\gamma$ with a fixed $\lambda=0.4$.}
	\label{fig:4}
\end{figure}

\textbf{Ablation Study}
We further conduct ablation study to investigate the key components and hyperparameters of FedSODA. The results prove the efficacy of each module in our framework in Table \ref{tab:t2}. Our approach effectively mitigates the adverse impact of data heterogeneity, thereby enhancing the model's dynamic aggregation capabilities. Additionally, we further analyze how the weighted degree $\lambda$ and attenuation coefficient $\gamma$ affect the performance of our method. Excessive retention of local information may hinder clients from identifying shared features during aggregation, whereas constrained retention contributes to the stability of local model optimization, especially in the presence of significant heterogeneity. In Fig. \ref{fig:4} a), our FedSODA reaches the highest Dice with $\lambda = 0.4$, while in Fig. \ref{fig:4} b), retaining 25\% of historical information yielded the best memory bank performance.

\begin{table}
	\centering
	\scalebox{0.8}{
		\begin{tabular}{ccc|c}
			\hline
			\textbf{SO} & \textbf{DA} & \textbf{$\mathcal{L}_{sc}$} & \textbf{Average Dice $\%$} \\ \hline
			&  &  & 82.62 \\ \hline
			$\checkmark$&  &  & 82.80 \\ \hline
			&$\checkmark$  &  & 82.97 \\ \hline
			$\checkmark$ & $\checkmark$ &  & 83.11 \\ \hline
			$\checkmark$ & $\checkmark$ & $\checkmark$ & \textbf{83.41} \\ \hline
		\end{tabular}
	}
	\caption{The ablation study of each module in FedSODA.}
	\label{tab:t2}
\end{table}

\section{Conclusion}
\label{sec:typestyle}
In this paper, we focus on the histopathology nuclei and tissue segmentation task across multiple sites. FedSODA is the first federated learning framework that makes effective use of the most extensive datasets for histopathology nuclei and tissue segmentation. We propose a synthetic-driven cross-assessment operation to effectively mitigate the sample imbalance across clients and a dynamic stratified-layer aggregation strategy to enhance shared representation and eliminate heterogeneity. We conduct extensive experiments on seven histopathology datasets and demonstrate the effectiveness of our method. Overall, our work contributes to fostering a broader impact of FL in real-world medical applications.

%

\bibliographystyle{IEEEbib}
\bibliography{Template}

\end{document}